\documentclass[
    ,final            
  ,draft            
  ]
  {aipproc}

\layoutstyle{6x9}
\newcommand{\lsim}{\raisebox{-0.7ex}{$\stackrel{\textstyle <}{\sim}$ }}
\newcommand{\vlk}{V_{{\rm low}\,k}}
\newcommand{\la}{\Lambda}
\newcommand{\fmi}{\, {\rm fm}^{-1}}
\begin{document}

\title{Nuclei from QCD : Strategy, Challenges and Status}



\author{Martin J. Savage}{
  address={Department of Physics, University of Washington, Seattle, WA 98195-1560.}
}

\begin{abstract}
I describe progress that is being made toward calculating the
properties and interactions of nuclei from QCD.
\end{abstract}

\maketitle

\section{Introduction}

Perhaps the greatest challenge facing those of us working in the area of strong
interaction physics is to be able to rigorously
compute the properties and interactions of nuclei.
The many decades of theoretical and experimental investigations in nuclear
physics have, in many instances, provided a very precise phenomenology of the
strong interactions in the non-perturbative regime.  However, at this point in
time we have little understanding of much of this phenomenology in terms of the underlying
theory of the strong interactions, Quantum Chromo Dynamics (QCD).
I wish to discuss a strategy for making a connection
between QCD and nuclear physics, which ultimately will allow for the calculation of nuclear
properties and processes in terms of the light quark
masses, the scale of the strong interactions, and the electroweak couplings.

\section{QCD to Nuclei: Strategy}

The ultimate goal is to be able to rigorously compute the properties and
interactions of nuclei from QCD.  This includes determining how the structure
of nuclei depend upon the fundamental constants of nature. 
Perhaps as important,
we would then  be in the position to reliably
compute quantities that cannot be accessed, either directly or indirectly,
by experiment.

The only way to rigorously compute strong-interaction quantities in the
nonperturbative regime is with lattice QCD.  One starts with the QCD Lagrange
density and performs a Monte-Carlo evaluation of Euclidean space Green
functions directly from the path integral.  
To perform such an evaluation, space-time is latticized and computations are
performed in a finite volume, at finite lattice spacing, and at this point in
time, with quark masses that are larger than the physical quark masses.
To compute any given quantity, contractions are performed in which the
valence quarks that propagate on any given gauge-field configuration are ``tied
together''.  For simple processes such as nucleon-nucleon scattering, such
contractions do not require significant computer time compared with lattice or propagator
generation.  However, as
one explores processes involving more hadrons, the number of contractions grows rapidly
(for a nucleus with atomic number $A$ and charge $Z$, the number of
contractions is $(A+Z)!(2A-Z)!$),
and a direct lattice QCD calculation of the properties of a large nucleus 
is quite impractical simply due to the computational time required.

The way to proceed is to establish a small number of effective theories, each
of which have well-defined expansion parameters and can be shown to be the most
general form consistent with the symmetries of QCD.
Each theory must provide a complete description of nuclei over some range of atomic
number.  Calculations in two ``adjacent'' theories are performed for a range of
atomic numbers for which both theories converge.  One then matches coefficients
in one EFT to the calculations in the other EFT or to the lattice, and thereby one can make an
indirect, but rigorous connection between QCD and nuclei.
It appears that four different matchings are required:
\begin{enumerate}
\item 
{\bf Lattice QCD}.
Lattice QCD calculations of the properties of the very
lightest nuclei will be possible at some point in 
the not so distant future~\cite{Savage:2005ma}.
Calculations for $A\le 4$ as a function of the light-quark masses, would
uniquely define the interactions between nucleons up to and including the four-body
operators.
Depending on the desired precision, one could
possibly imagine calculations up to $A\sim 8$.
\item
{\bf Exact Many-Body Methods}.
During the past decade one has seen remarkable progress in the calculation of
nuclear properties using Green Function Monte-Carlo (GFMC) with the
$AV_{18}$-potential (e.g. Ref.~\cite{Pieper:2004qw}) 
and also the No-Core Shell Model (NCSM) (e.g. Ref.~\cite{Forssen:2004dk})
using chiral potentials.
Starting with the chiral potentials, which are the most general interactions
between nucleons consistent with QCD, one would calculate the properties of
nuclei as a function of all the parameters in the chiral potentials 
with GFMC or the NCSM
out to some given order in the chiral expansion.  A comparison between such calculations
and lattice QCD calculations will determine these parameters to some level
of precision.  These parameters can then be used in the calculation of nuclear
properties up to atomic numbers $A\sim 20-30$.
The computer time for these many-body theories suffers from the same $\sim
(A!)^2$ blow-up that lattice QCD does, and for a sufficiently large nucleus,
such calculations become impractical.

Another recent development that shows exceptional promise is the latticization
of the chiral effective field 
theories~\cite{Muller:1998rc,Muller:1999cp,Lee:2004si,Borasoy:2005yc,Seki:2005ns}.
This should provide  a model-independent calculation of nuclear processes once
matched to lattice QCD calculations.
\item
{\bf Coupled Cluster Calculations}.
In order to move to larger nuclei, $A\lsim 100$ a technique that has shown
promise is to implement a coupled-clusters expansion (e.g. Ref.~\cite{Wloch:2005qq}).
One uses the same chiral potential that will have been matched to lattice QCD
calculations, and then performs a diagonalization of the nuclear Hamiltonian,
after truncating the cluster expansion, which itself contains arbitrary coefficients.
The results of these calculations will be matched to those of the NCSM or GFMC
for $A\sim 20-30$
to determine the arbitrary coefficients.
This method is unlikely to be practical for very large atomic numbers.
\item
{\bf Density Functional Theory (??) and Very Large Nuclei}
To complete the periodic table one needs to have an effective theory that is
valid for very large nuclei and nuclear matter.
A candidate that has received recent attention is Density Function Theory
(DFT) (e.g. Refs.~\cite{Furnstahl:2004xn,Schwenk:2004hm}).  
It remains to be seen if this is in fact a viable candidate.
There is reason to hope that this will be useful because 
there is clearly a
density expansion in large nuclei with a power-counting that is consistent with
the 
Naive Dimensional Analysis (NDA) of Georgi and Manohar~\cite{Manohar:1983md}.
The application of DFT to large nuclei is presently the
least rigorously developed component of this program.

The latticized chiral theory mentioned previously can also be applied to the
infinite nuclear matter problem.  This work is still in the very earliest stages
of exploration, but this looks promising~\cite{Lee:2004si}.
\end{enumerate}

\section{QCD to Nuclei: One of the Challenges}

An intriguing aspect of nuclear physics and QCD that has
slowed the theoretical progress in  connecting QCD to nuclear physics
is the fine-tunings that are present.
I will discuss just two of these fine-tunings.

\subsection{$3\alpha\rightarrow ^{12}C$}

Perhaps the most famous fine-tuning is that observed in the triple-$\alpha$ process.
The production of carbon in stars results
from the reactions $3\alpha \leftrightarrow \alpha + ^8Be^* \leftrightarrow ^{12}~C^{**}$
being in thermal equilibrium.
Because the ground state of $^8Be$ is barely unbound and the second
excited state in $^{12}C$ is where it is,
these reactions can simultaneously be in thermal equilibrium at temperatures $\sim T_8$.
Further, the state in $^{16}O$ that could potentially be
populated via $\alpha+ ^{12}C$ is sub-threshold, and there is a large energy
splitting to the next state in  $^{16}O$,
preventing significant carbon destruction.
Much has been made about the positions of these levels, 
and in fact the location of the $^{12}C^{**}$
was predicted  prior to its discovery based upon anthropic arguments.
Of the many possible universes with random values of the  fundamental
constants, as might arise from the landscape~\cite{Kachru:2003aw,Susskind:2003kw} 
scenario in string theory,
sufficient $^{12}C$ will be
produced to support carbon-based life
only in those universes with energy levels in the $A=12$
system that are very close to those observed.

As a  first step toward understanding these fine-tunings, there has been recent
work in which limits have been placed on the variation in the magnitude of the
nucleon-nucleon (NN) potential that is consistent with the production of 
{\it significant} amounts of $^{12}C$.  It was found that a change of $\sim
0.5\%$ in the strength of the NN interaction was sufficient to yield a universe
that does not contain significant amounts of $^{12}C$ or
$^{16}O$~\cite{Oberhummer:2000zj,Csoto:2000iw}.
There has also been recent work exploring the dependence of 
 $^{12}C$ and  $^{16}O$ abundances upon the location of the $^{12}C^{**}$ 
level~\cite{Schlattl:2003dy}.

What is at the heart of these fine-tunings is not so much the absolute location
of the energy-levels, but their relative location.
It is unlikely that the simplest variations that one can imagine, 
changing the energy of only the $^{12}C^{**}$ level and determining abundances,
actually provide an indication of how robust this system is.  
It would be a wonderful accomplishment to explore every aspect of
these systems, and these fine-tunings in terms of the fundamental parameters of
nature, the light-quark masses $m_q$, the scale of the strong interaction
$\Lambda_{\rm QCD}$, and the electromagnetic coupling $\alpha_e$.
However, at present we are far from being able to perform such a
study due to both a lack of computational power, and a lack of theoretical
infrastructure.
Only crude estimates of how nuclear properties and
interactions depend upon the fundamental constants are 
possible~\cite{Flambaum:2002de,Flambaum:2002wq}.

When considered in terms of QCD, as opposed to nuclear structure,
the fine-tunings in this system are quite severe.
The location of the $i^{th}$ energy
level is of the form
\begin{eqnarray}
E_i & = & \Lambda_{\rm QCD} \ f_i ( {m_u\over\Lambda_{\rm QCD}}, 
{m_d\over\Lambda_{\rm QCD}}, {m_s\over\Lambda_{\rm QCD}}, \alpha_e)
\ \ \ ,
\label{eq:c12levels}
\end{eqnarray}
and given that the scale of strong interactions is hundreds of MeV, and the allowed
variation in the relative location of the level in $^{12}C$ is $\sim 100~{\rm
  keV}$, there is a fine-tuning between the $f_i$ at the level of $10^{-4}$.

\subsection{Nucleon-Nucleon Interactions}

The NN interaction itself is finely-tuned.
The NN potential can be roughly separated into three distance-scales, the
long-range part, the intermediate range part and the short-distance part.
The long-range part is unambiguously described by one-pion-exchange (OPE), both
theoretically and also by fitting to the multitude of scattering data.
The intermediate range interaction (attraction), which traditionally was considered to
result from the exchange of a ``$\sigma$-meson'', has recently been shown to be
the result of two-pion exchange (TPE)~\cite{Rentmeester:2003mf} as calculated using chiral perturbation
theory ($\chi$PT).  There is no reason to believe that the  short-range component of the potential is
describable in terms of meson exchanges, and the ``best'' potentials
(defined by the value of $\chi^2$ in fitting) have some short-distance functional form consistent
with power-counting expectations from effective field theory (EFT).
The typical distance scale of the long-distance component is $\sim 1/m_\pi$, of
the intermediate range component is $\sim 1/(2 m_\pi)$, and of the
short-distance component is 
$\sim 1/m_\rho$.  
The S-wave NN wavefunctions emerging from these potentials are essentially
horizontal, which is highly unnatural and requires a fine-tuning between the
various component of the potential.  
The deuteron has a binding energy of $\sim 2.2~{\rm MeV}$, and the 
scattering length in the $^1 S_0$-channel is $\sim -24~{\rm fm}$.
The EFT describing the NN 
interaction~\cite{Weinberg:1990rz,Weinberg:1991um,Ordonez:1995rz,Kaplan:1998we,Kaplan:1998tg,Beane:2001bc} 
is considerably different to EFT's that
one is familiar with.  For most EFT's one can count the dimension of an
operator and determine the size of its contribution to a process. This is not
true for the EFT describing NN interactions, as one has to perform an expansion
about a non-trivial, unstable infrared fixed point in the renormalization group
(RG) 
flow~\cite{Kaplan:1998we,Kaplan:1998tg,Birse:1998dk,Birse:1998tm}.
The implication of this is that the dimension-6 four-nucleon operator 
contribution to NN-scattering 
is not suppressed compared to that from the dimension-4 pion-nucleon
interaction.
In fact, in the $^1 S_0$ channel, the long-distance 
pionic effects can be treated as a  
perturbation~\cite{Kaplan:1998we,Kaplan:1998tg,Savage:1998vh} and the full utility of the RG is
explicit.

\section{QCD to Nuclei: Status}

During the past few years there has been substantial progress toward being able
to compute nuclear properties from QCD using the strategy already outlined.  

\subsection{Lattice QCD}

Lattice QCD has entered an era in which reliable calculations of strong
interaction quantities can be performed with fully-dynamical QCD calculations
at small lattice spacings and in large volumes (large and small are defined
relative to the scale of chiral symmetry breaking).  The lattice actions have
good chiral symmetry through the invention of Domain-Wall
fermions~\cite{Kaplan:1992bt,Shamir:1993zy} 
and Overlap fermions~\cite{Narayanan:1993sk}.  
Further, there has been substantial progress in chiral EFT's 
that, in addition to describing the light-quark mass dependence and allowing
for rigorous chiral extrapolations, 
facilitate the  removal of finite-lattice
spacing~\cite{Rupak:2002sm} and finite-volume effects inherent in the lattice
QCD calculations, e.g. Refs.~\cite{Colangelo:2003hf,Beane:2004tw}.

\begin{figure}[!ht]
  \includegraphics[height=0.25\textheight]{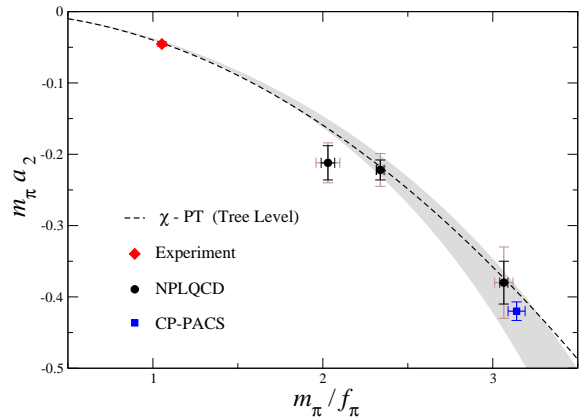}
  \caption{$I=2$ $\pi\pi$ scattering from fully-dynamical lattice 
QCD~\protect\cite{Beane:2005rj}.
(This figure is taken from Ref.~\protect\cite{Beane:2005rj}.)}
\label{fig:pipi}
\end{figure}

There has been much effort over the years to precisely determine strong
interaction matrix elements required to extract parameters of the
electroweak theory, such as $V_{bc}$.  
A subset of these were described in Chris Sachrajda's
talk~\cite{SachrajdaPanic05}, 
and I will not
discuss them here.
A recent calculation involving the light mesons of interest to nuclear
physicists is the calculation of $I=2$ $\pi\pi$ scattering in fully dynamical
QCD by the NPLQCD collaboration~\cite{Beane:2005rj},
as shown in fig.~\ref{fig:pipi}.
One finds good agreement with the predictions of chiral perturbation theory,
and the calculations are at small enough pion masses where the perturbative
expansion is reliable (see also Ref.~\cite{Chen:2005ab}).
\vskip 0.3in
Compared to the
meson sector, there has been somewhat less emphasis on the baryon sector.  
However, this is changing through the significant investment in
lattice QCD at the Jefferson Laboratory by nuclear physics DOE and SciDAC.   
There are several hundred processors available for lattice calculations, and 
more importantly is the work of Robert Edwards and his team to develop and make
available the lattice software suite {\it Chroma}~\cite{Edwards:2004sx,sse2}.

\begin{figure}[!ht]
  \includegraphics[height=0.35\textheight,angle=-90]{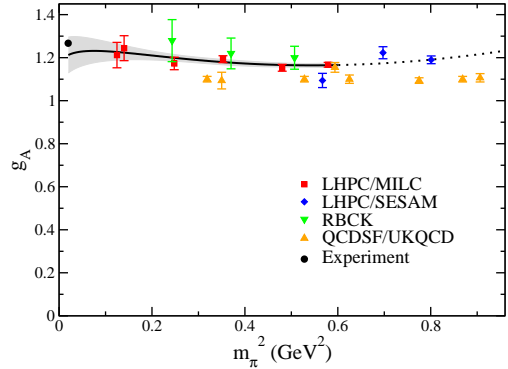}
  \caption{The light-quark isovector axial current matrix element in the
    nucleon computed in fully dynamical lattice
    QCD~\protect\cite{Edwards:2005ym}, and its chiral extrapolation.
(This figure is taken from Ref.~\protect\cite{Edwards:2005ym}.)
}
\label{fig:gA}
\end{figure}

\begin{figure}[!ht]
  \includegraphics[height=0.23\textheight]{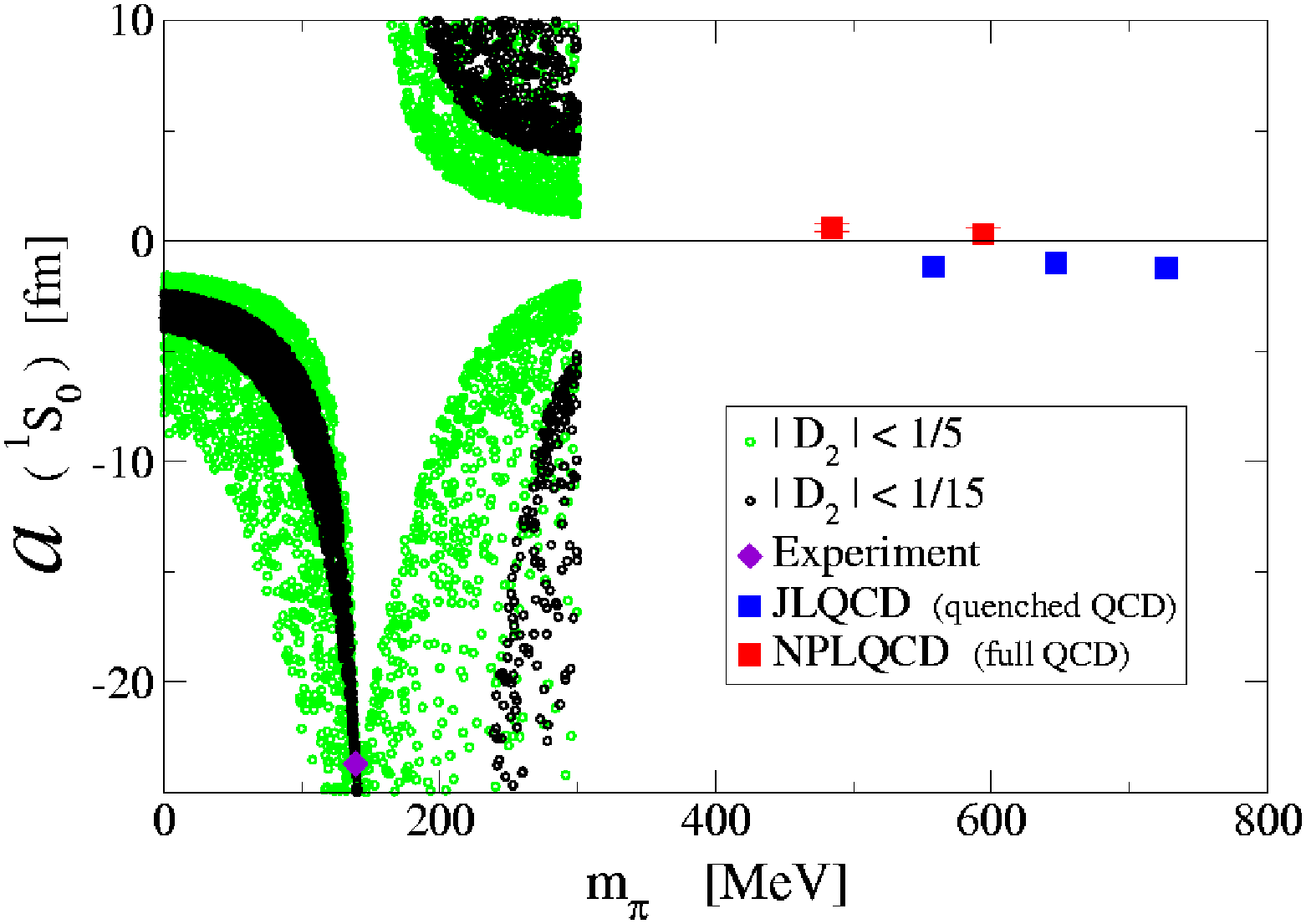}\qquad
  \includegraphics[height=0.23\textheight]{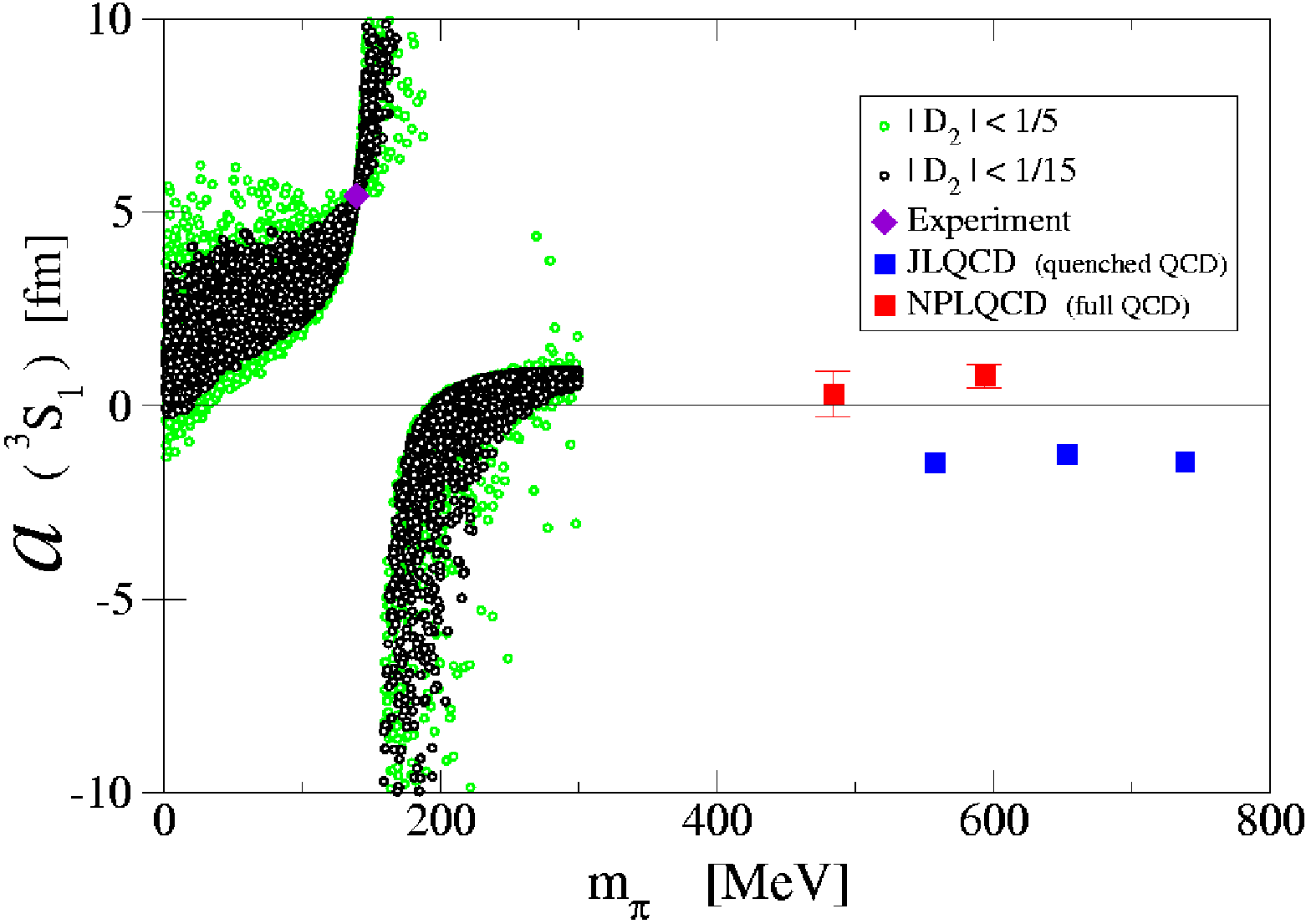}
  \caption{The nucleon-nucleon scattering lengths in the $^1S_0$ channel (left
    panel) and the $^3S_1-^3D_1$ coupled channels (right panel) as a function
    of the pion mass.  The light (green) and dark (black) sets of points denote present
theoretical estimates of the quark-mass dependence of the scattering lengths
based upon EFT arguments~\protect\cite{Beane:2002xf}.
The QCD data points at $m_\pi \sim 500~{\rm MeV}$ and $\sim 600~{\rm MeV}$ 
are the {\bf preliminary} results of the NPLQCD exploratory investigation, while the other data points
are the results of a quenched calculation~\protect\cite{Fukugita:1994ve}.
}
\label{fig:NN}
\end{figure}

There has been very impressive recent work by LHPC~\cite{Edwards:2005ym} computing 
the matrix element of the light-quark axial current in the nucleon at low quark
masses and large volumes, as shown in fig.~\ref{fig:gA}.
Further, there are some preliminary results from the NPLQCD collaboration for
the nucleon-nucleon scattering lengths in fully-dynamical lattice QCD, as shown
in  fig.~\ref{fig:NN}.

\subsection{Light Nuclei, Chiral Symmetry and the Renormalization Group}

There has been impressive progress in the calculation of  properties of nuclei
when the NN, 3N and 4N interactions are specified.
Calculations using GFMC have developed to the stage where, in addition to the
ground states,
the excited states of the light nuclei can be extracted~\cite{Pieper:2004qw}.
\begin{figure}[!ht]
  \includegraphics[height=0.35\textheight,angle=-90]{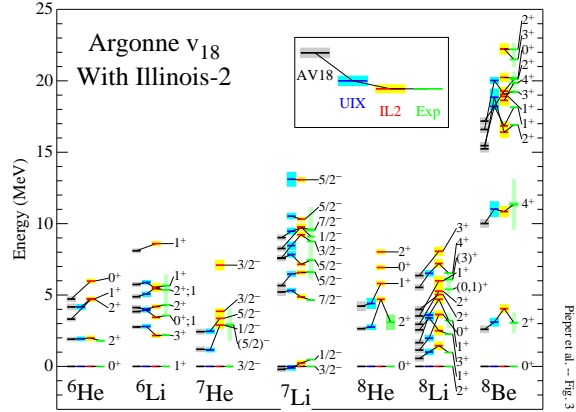}
  \caption{The spectra of the $A=6,7,8$ nuclei computed with a GFMC from the 
$AV_{18}$ and IL2 interactions~\protect\cite{Pieper:2004qw}.
(This figure is taken from Ref.~\protect\cite{Pieper:2004qw}.)
}
\label{fig:gfmc}
\end{figure}
The agreement between the calculated energy-levels and those observed is truly
impressive,
and clearly demonstrates the strength of this technique.
An example of this agreement can be seen in fig.~\ref{fig:gfmc}.
One would like to see these calculations performed with chiral potentials, so
that they could be matched to lattice QCD calculations of the future.

\begin{figure}[!ht]
  \includegraphics[height=0.4\textheight]{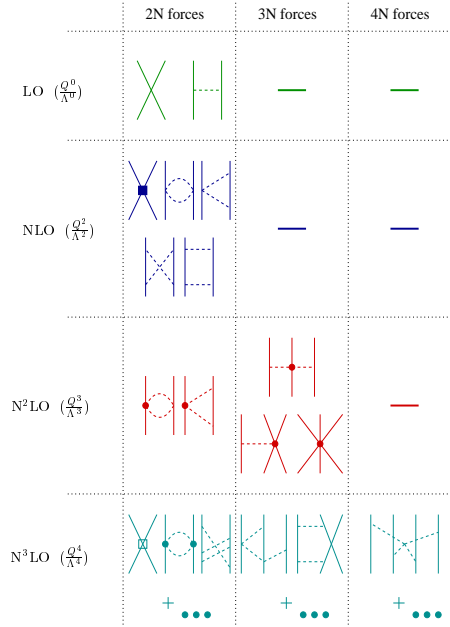}
  \caption{Interactions between nucleons as classified in Weinbergs'
    power-counting scheme~\protect\cite{Weinberg:1990rz,Weinberg:1991um}.
The solid lines denote nucleons, while the dashed lines denote pions.
(This figure is taken from Ref.~\protect\cite{Gloeckle:2003ub}.)
}
\label{fig:Weinberg}
\end{figure}
Classifying and computing the interactions between nucleons based upon the 
approximate chiral symmetry of
QCD~\cite{Weinberg:1990rz,Weinberg:1991um,Ordonez:1995rz,Kaplan:1998we,Kaplan:1998tg,Beane:2001bc}
is now at an advanced stage of development, e.g. 
Refs.~\cite{Gloeckle:2003ub,Epelbaum:2004fk,Nogga:2005hy,Platter:2005sj}.
Initiated by the pioneering papers of Weinberg in the early 1990's, 
the field is currently at the stage of having determined the small expansion
parameter and to have essentially determined (in terms of {\it apriori} unknown
counterterms)
the interactions between two, three and four nucleons out to four orders in the
expansion~\cite{Epelbaum:2005zc}, see fig.~\ref{fig:Weinberg}.
The importance of this effort cannot be overstated.  In order to make rigorous,
model-independent predictions and calculations in nuclear physics, the most
general form of the interactions consistent with QCD must be known.  
The established power-counting finds that contributions from operators
involving four or more nucleons are parametrically suppressed.
In addition to establishing a rigorous framework, the
light-quark mass dependence of nuclear interactions is provided by these same
interactions.

\begin{figure}[!ht]
  \includegraphics[height=0.2\textheight]{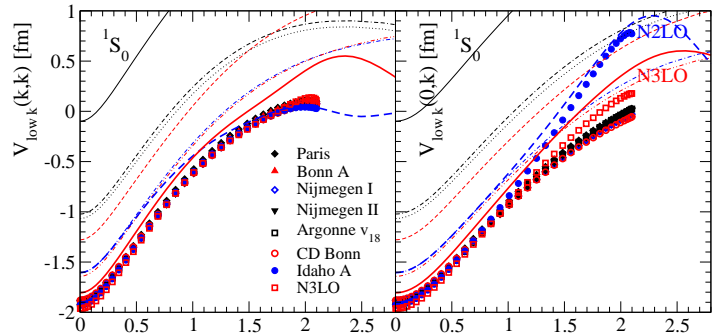}
  \caption{Diagonal (left) and off-diagonal (right)
momentum-space matrix element for $\vlk$ (symbols)  in the $^1S_0$ channel
versus relative
momentum derived from different modern potential models
for $\la = 2.1 \fmi$~\protect\cite{Schwenk:2004hz}. 
The bare interactions are shown 
as lines, while the thick sold and thick dashed lines denote the N2LO (Idaho A)
or N3LO interactions, resp.
(This figure is taken from Ref.~\protect\cite{Schwenk:2004hz}.)
}
\label{fig:vlowk}
\end{figure}

The RG is a valuable tool for studying quantum systems and
has been employed by particle physicists for decades.
In the course of developing the EFT's for nuclear physics, it was shown that
the RG is also a powerful tool for nuclear 
physics~\cite{Kaplan:1998we,Kaplan:1998tg,Beane:2001bc,Birse:1998dk,Birse:1998tm}.
Important nuclear physics phenomenology has arisen from applying the RG
framework to the modern phenomenological NN interactions, such as the $AV_{18}$
potential, CD-Bonn potential and Idaho A potential.  It was shown that by
evolving these potential down to a sufficient low scale,
$\Lambda\sim 600~{\rm MeV}$, they all coincide in
momentum-space~\cite{Bogner:2001gq,Bogner:2001jn} 
to what is now referred to as $\vlk$,
(for a nice overview see 
Ref.~\cite{Schwenk:2004hz}), as shown for the $^1S_0$ channel in fig.~\ref{fig:vlowk}.
The success of the chiral EFT program for these interactions
meant that this result had to be true.
As the renormalization scale of  $\vlk$
is lower than the typical scale of the hard-core
interaction in the ``bare''-potentials, the interactions are softer, and as a result
the many-body calculations in  nuclei are significantly 
more convergent than 
with the bare potentials.

\subsection{The No-Core Shell Model}

\begin{figure}[!ht]
  \includegraphics[height=0.26\textheight]{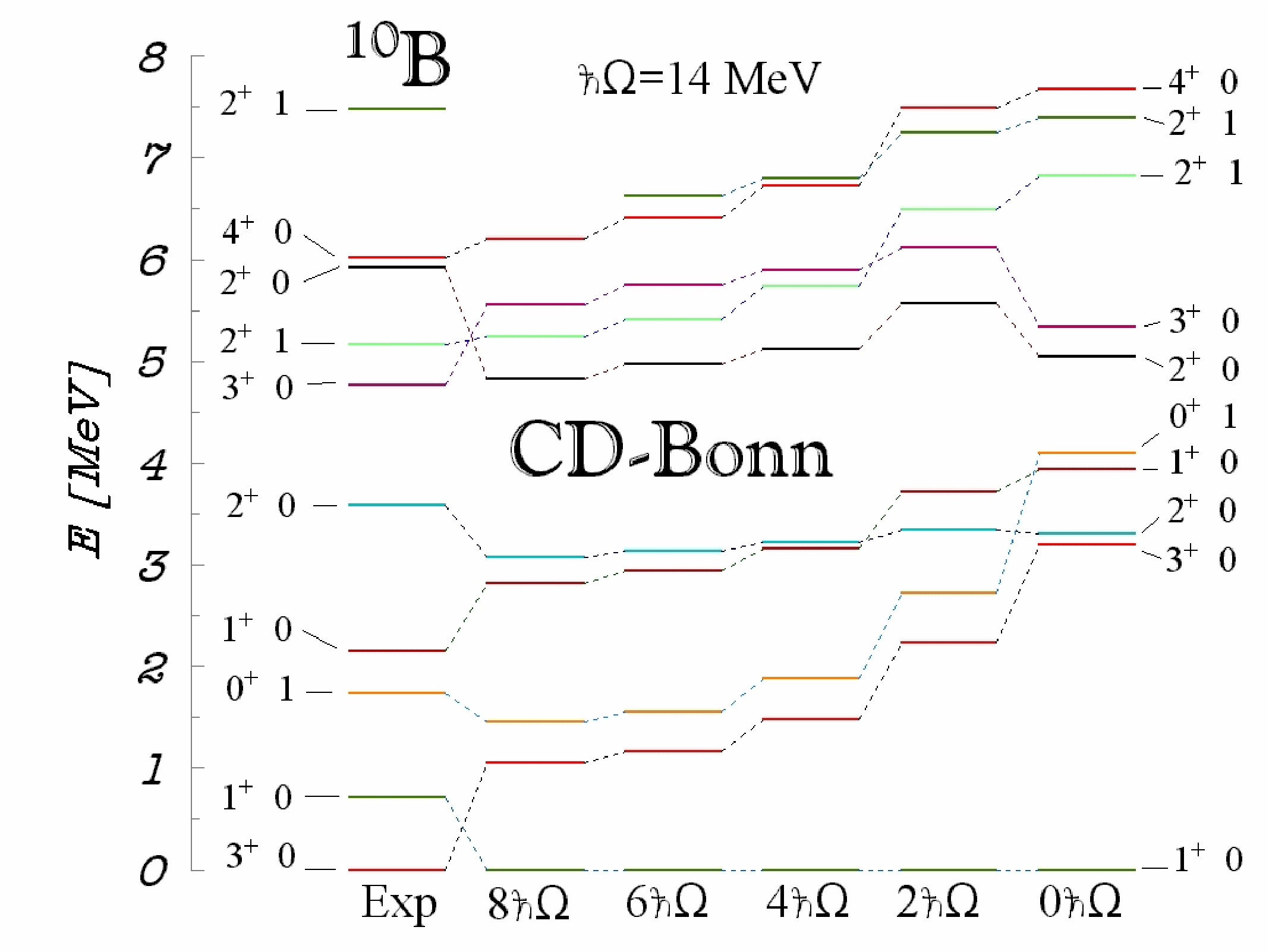}\ \ 
  \includegraphics[height=0.26\textheight]{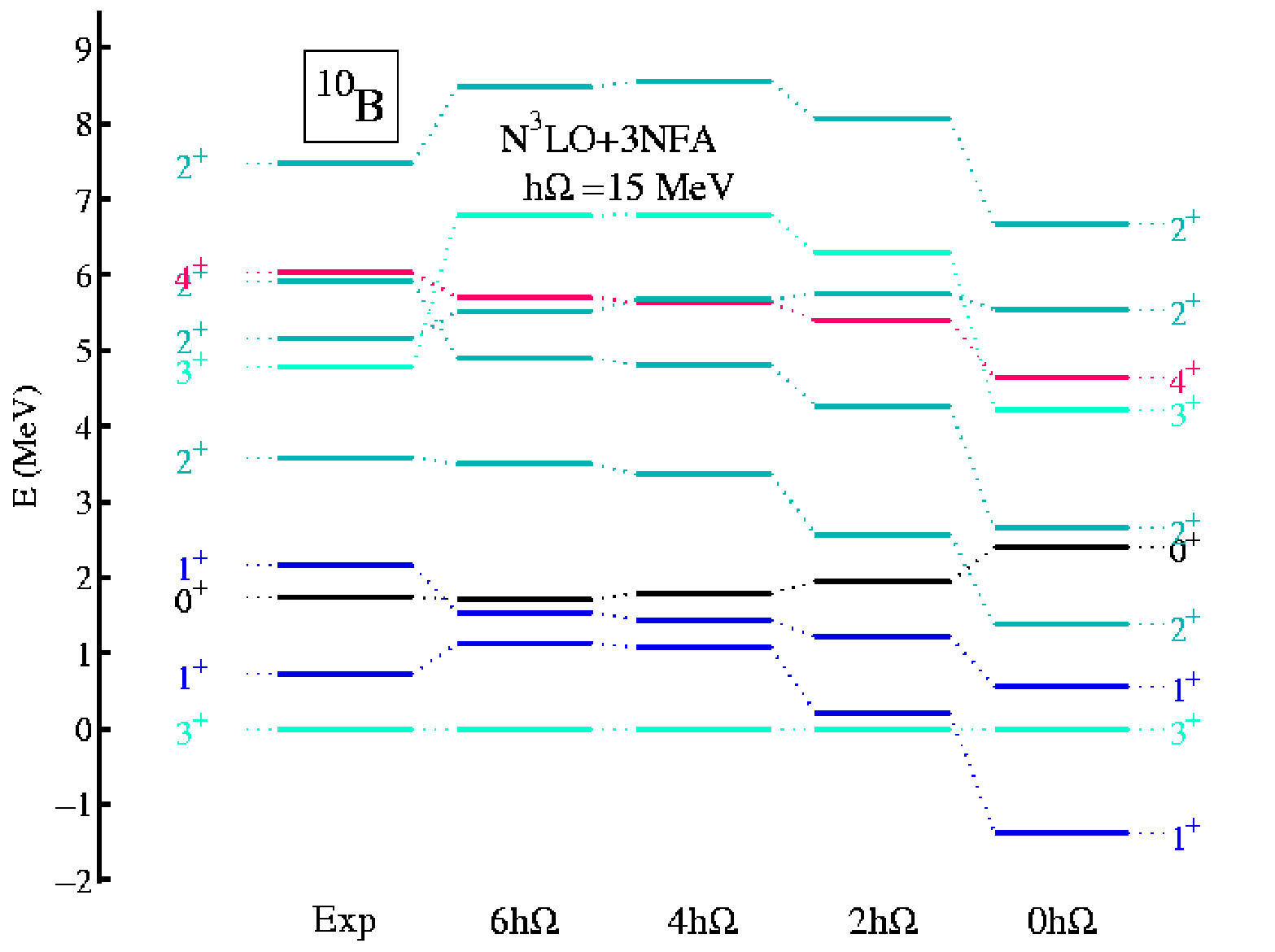}
  \caption{The spectrum of $^{10}B$, determined in the NCSM with the CD-Bonn
    potential (left panel) and the chiral potential (right panel) outlined in
    the text, as a function of the size of the basis.
}
\label{fig:Bten}
\end{figure}

A significant step toward the rigorous calculation of nuclear properties is the
development of the NCSM.  The entire goal of this program is to implement
effective interaction theory, and ``take the model out of the shell model''.
Historically, the shell model implies that there is a small number of
``active'' nucleons in shell model orbits outside an inert core of nucleons.
The Hamiltonian of the active space is diagonalized to yield the energy
eigenstates and energies.
The NCSM treats all nucleons as active particles.  Some arbitrary but complete 
basis is chosen,
conventionally that of a harmonic oscillator (HO), and the Hamiltonian
is constructed in this basis for realistic NN, $3N$ and recently $4N$
interactions.
If an infinite number of HO states
were included in the calculation,  the eigenstates 
and energies would be independent of
the HO parameter.
This is impractical, but as the model space in enlarged the eigenenergies and
states become less dependent upon the scale of the HO.
A sufficient number
of HO levels can be included to obtain the desired precision 
(for a discussion see Ref.~\cite{Beane:2000fx}).
Recent calculations 
with the softer potentials that result from the chiral EFT's or $\vlk$ are very
encouraging.
In particular, calculations have been done for $A=10$
with the NN interaction out to NNNLO, the 3-body interaction out to NNLO and
fit to the properties of $A=3$ and $A=4$ nuclei.
Fig.~\ref{fig:Bten} shows the energy levels of $^{10}B$ computed in the NCSM with
the CD-Bonn potential, and with the chiral 
potential~\footnote{I would like to thank Erich Ormand and his collaborators
for allowing me to show these figures.}.
Clearly, the convergence is greatly improved when the chiral potential is used.
\vskip 0.1in

\subsection{The Braaten-Hammer Conjecture}

\begin{figure}[!ht]
  \includegraphics[height=0.2\textheight]{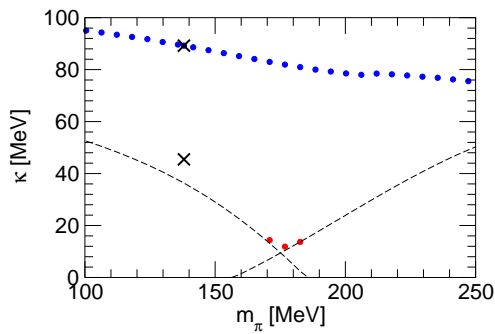}
  \caption{The binding momenta $\kappa=(mB_3)^{1/2}$ 
of $p n n$ bound states as a function of the pion mass.
The circles indicate the triton ground state and excited state. 
The crosses give the binding energy of the physical
deuteron and triton, while the dashed lines give the thresholds
for decay into a nucleon plus a deuteron (left curve) or 
a spin-singlet di-nucleon (right curve).
(This figure is taken from Ref.~\protect\cite{Braaten:2003eu}.)
}
\label{fig:triton}
\end{figure}

An interesting conjecture was put forth a couple of years ago 
by Braaten and Hammer~\cite{Braaten:2003eu}.
As discussed earlier, one has a rough idea of the quark mass dependence of the 
NN sector~\cite{Beane:2002xf,Beane:2002vs,Epelbaum:2002gb}, 
see fig.~\ref{fig:NN}.
As the pion mass is increased, it is possible
that the scattering lengths in both the $^1S_0$ and $^3S_1-^3D_1$ channels
become  large.  In fact, considering the limits shown in
fig.~\ref{fig:NN}, when the pion mass is around $\sim 175~{\rm
  MeV}$ an additional shallow bound state appears in the spectrum of the
triton,
as shown in fig.~\ref{fig:triton}.
This prediction is something that could be explored with lattice QCD, 
once the formalism is put in place for dealing with 3-body
systems in Euclidean space at finite-volume.

Somewhat more tantalizing, and also something that could be explored with
lattice QCD, is their conjecture that the up and down quark masses could be
individually tuned to values for which the scattering lengths in both the 
$^1S_0$ and $^3S_1-^3D_1$ channels are infinite.
In such a scenario, the system is invariant under discrete scale
transformations toward the infrared~\cite{Bedaque:1998kg},
and the triton has an infinite number of bound states,
with the energy of adjacent states related by
$E_{n+1}^2 = 515\  E_n^2$.
\vskip 0.5in
%

\section{Summary and Outlook}

Important progress has been made in the areas necessary for 
calculation of the properties and interactions of nuclei
from QCD.
It appears that we are entering an era in which 
lattice QCD calculations in the 
$A=2,3,4$ systems will be matched onto the few-nucleon chiral interactions.
These interactions will then be used to compute the properties of nuclei via 
GFMC, the NCSM, or potentially the latticized chiral theory.

\begin{theacknowledgments}
I would like to thank  S. Beane, P. Bedaque, K. Orginos, E. Ormand
and A. Schwenk for discussions.
This work is supported in part by the U.S.~Dept.~of
Energy under Grant No.~DE-FG03-97ER4014.
University of Washington preprint number : NT@UW-05-016.

\end{theacknowledgments}

\end{document}